\documentclass[letterpaper, 10 pt, conference]{ieeeconf}  

\usepackage{amsmath,amsfonts,amsthm}
\usepackage{graphicx}
\usepackage[strict=true,style=english]{csquotes}
\usepackage{caption}
\usepackage{subcaption}
\usepackage{xcolor}
\usepackage{comment}
\usepackage{cite}

\theoremstyle{definition}
\newtheorem{Problem}{Problem}

\newtheorem{Remark}{Remark}

\newtheorem{Case Study}{Case Study}
\usepackage[linesnumbered,ruled,vlined]{algorithm2e}

\IEEEoverridecommandlockouts                              

\title{\LARGE \bf
Safe Model-based Control from Signal Temporal Logic Specifications Using Recurrent Neural Networks
}

\author{Wenliang Liu$^{1}$, Mirai Nishioka$^{2}$, Calin Belta$^{1}$
\thanks{This work was partially supported by the National Science Foundation under grants IIS-2024606 and IIS-1723995.}
\thanks{$^{1}$Wenliang Liu and Calin Belta are with Boston University, Boston, MA, USA
        {\tt\small wliu97@bu.edu, cbelta@bu.edu}}
\thanks{$^{2}$Mirai Nishioka is with Commonwealth School, Boston, MA, USA
        {\tt\small mnishioka@commschool.org}}}

\begin{document}

\maketitle
\thispagestyle{empty}
\pagestyle{empty}

\begin{abstract}
We propose a policy search approach to learn controllers from specifications given as Signal Temporal Logic (STL) formulae. The system model, which is unknown but assumed to be an affine control system, is learned together with the control policy. The model is implemented as two feedforward neural networks (FNNs) - one for the drift, and one for the control directions. To capture the history dependency of STL specifications, we use a recurrent neural network (RNN) to implement the control policy. In contrast to prevalent model-free methods, the learning approach proposed here takes advantage of the learned model and is more efficient. We use control barrier functions (CBFs) with the learned model to improve the safety of the system. We validate our algorithm via simulations and experiments. The results show that our approach can satisfy the given specification within very few system runs, and can be used for on-line control. 
\end{abstract}

\section{INTRODUCTION}
\label{sec:intro}

Temporal logics, such as Linear Temporal Logic (LTL) and Computation Tree Logic (CTL) \cite{baier2008principles}, have been increasingly used as specification languages in robotics, biology, and autonomous driving applications. 
In this paper, we use Signal Temporal Logic (STL) \cite{maler2004monitoring}, which was originally developed to monitor temporal properties of real-valued signals. STL is equipped with both qualitative semantics, which measures whether a signal satisfies a formula, and quantitative semantics, known as robustness, which assigns a real value to measure how strongly a signal satisfies a formula \cite{donze2010robust}. Higher robustness scores indicate stronger satisfactions. Using robustness, controlling a system from an STL specification can be formulated as an optimization problem with robustness as a constraint, as a term in the objective function, or both. Such a problem can be solved using either Mixed Integer Programming (MIP) \cite{raman2014model}, \cite{sadraddini2015robust}, or gradient-based methods \cite{pant2017smooth, haghighi2019control,varnai2020robustness,gilpin2020smooth}. However, both MIP and gradient-based methods are computationally expensive, especially for high dimensional systems and long-horizon planning, which prevents real-time implementation. 

Neural Network (NN) - based controllers have been proposed to address the computational complexity of the methods described above. NNs are trained offline and executed online, which enables real-time implementation. In \cite{yaghoubi2020worst}, the control policy is parameterized as a Feedforward Neural Network (FNN) to maximize the robustness of an STL formula defined over system trajectories. FNNs are memoryless controllers. However, the controller to satisfy an STL formula is, in general, \emph{history-dependent}. 
For instance, if a specification requires an agent to eventually visit two regions, the agent needs to know which region it has already visited so that it can go towards the other one. A Recurrent Neural Network (RNN) controller, which has memory, was proposed in \cite{liu2021recurrent} to satisfy STL specifications. The RNN was trained on a dataset containing satisfying trajectories. However, such a dataset is not always available. \cite{liu2021recurrent} used the optimization methods mentioned above to generate the dataset, which is again computationally expensive. 
The authors of \cite{leung2022semi} proposed a semi-supervised approach to train an RNN controller such that the resulting trajectory maximizes the STL robustness as well as minimizes the distance from a given dataset of expert demonstrations. Compared with \cite{liu2021recurrent}, the size of the dataset is largely reduced. All the above approaches assume the system dynamics are known, which is sometimes not the case in practice. RNNs were also used in sequential prediction tasks with STL constraints \cite{ma2020stlnet}, where neither system dynamics nor controls are considered. 

Reinforcement Learning (RL) can solve control problems for systems with unknown dynamics. Recently, RL approaches have been investigated to synthesize controllers satisfying STL specifications. The advantage of combining RL and STL is that the STL robustness can be used to generate RL rewards, which avoids reward hacking \cite{amodei2016concrete}. The authors of \cite{aksaray2016q} and \cite{venkataraman2020tractable} used Q-learning to learn control policies that maximize the robustness of STL fragments for systems with discrete state and control spaces. Similar ideas were used for continuous states and controls using deep RL \cite{balakrishnan2019structured,ikemoto2021deep}. These methods restrict the specification to be a fragment of STL. Path integral was used for learning control policies from Truncated Linear Temporal Logic (TLTL) \cite{li2018policy} and STL \cite{varnai2019prescribed}.
In \cite{li2019formal,wang2020continuous,cai2021modular}, automaton-based approaches were used with FNN controllers to satisfy LTL and TLTL specfications, where history information is contained in the automaton state. All the above methods are completely model-free, and are therefore data inefficient, i.e., they require a large number of trials to learn the policy 
\cite{deisenroth2011pilco}. 

Model-based policy search methods (e.g., \cite{deisenroth2011pilco}, \cite{gal2016improving}) learn a dynamic model of the system and use it to guide policy search. They are more data-efficient than model-free methods. 
In this paper, we propose a model-based policy search method that learns both the system model and the control policy from an STL specification. The model is unknown but assumed to be an affine control system. It is implemented as two FNNs with dropout (one for the drift and one for the control directions), which are approximations of Bayesian neural networks \cite{gal2016dropout} and can reduce model bias \cite{deisenroth2011pilco}.  The control policy is an RNN, which outputs the control depending on current state and past states, and therefore captures the memory necessary satisfy arbitrary STL formulas. Only a few system executions are needed for training the model, and no extra executions are needed to train the RNN controller. After training, the RNN controller can be executed very fast, which enables real time control. 

Another critical concern for RL is safety: unsafe states should never be reached during training and testing.
Control Barrier Functions (CBF) \cite{ames2019control} have been proposed to obtain safe control. In \cite{li2019formal}, CBFs are incorporated with formal methods to guarantee the safety of RL. Typically, using CBF requires the system dynamics to be known.
The authors of \cite{cai2021safety} use Gaussian process to model an uncertain system and use CBFs to guide a safe exploration.
Similarly, we consider an unknown system. We propose a method that applies CBF to the learned probabilistic model, which takes into account the model uncertainty. Applying CBF when training the model can also benefit the exploration, because without CBF, we have to often stop the system due to proximity to unsafe states. Another related work considering unknown dynamics is \cite{verginis2021non}, in which FNN controllers are trained, but an open-loop trajectory satisfying the STL specification needs to be computed before each run. 

The contributions of this paper are summarized as follows. (1) We propose a novel model-based policy search method to learn a policy that maximizes the expected robustness of a given STL specification with unknown system dynamics. Our approach considers full STL formulae, requires no dataset, and involves only few system executions. (2) We incorporate CBFs into the learning process with the consideration of model uncertainty, which improves the safety and exploration performance. (3) Simulations and experiments are implemented, illustrating the efficacy and scalability of our approach, and the ability of real-time control. 

\section{Notations and Preliminaries}

We consider a discrete-time affine control system:
\vspace{-2pt}
\begin{equation}
    \label{eq:dynamics}
    x_{t+1} = f(x_t)+g(x_t)u_t,\;t=0,1,\ldots
    \vspace{-2pt}
\end{equation}
where $x_t\in\mathcal X\subseteq\mathbb R^n$ is the system state at time $t$, $u_t\in\mathcal U\subseteq\mathbb R^m$ is the control input at time $t$, and $f:\mathcal{X}\rightarrow\mathcal{X}$,  $g:\mathcal{X}\rightarrow\mathbb{R}^{n\times m}$ are locally Lipschitz continuous functions modeling the drift and the control directions.
Let $x_{t_1:t_2}$ denote the partial state trajectory $x_{t_1}, x_{t_1+1}, \ldots, x_{t_2}$ with $t_2>t_1$. 

\subsection{Signal Temporal Logic (STL)}

An $n$-dimensional real-valued signal is denoted as $X=x_0x_1\ldots\ $, where $x_t\in\mathbb R^n$, $t\in\mathbb Z_{\geq0}$.  
The STL {\em syntax} \cite{maler2004monitoring} is defined as:
\vspace{-5pt}
\begin{equation}
\label{eq:syntax}
\varphi:=\top|\;\mu \; | \; \neg\varphi \; | \; \varphi_1\land\varphi_2 \; |   \;  \varphi_1 \mathbf{U}_{I} \varphi_2,
\vspace{-3pt}
\end{equation}
where $\varphi$, $\varphi_1$, $\varphi_2$ are STL formulae, $\top$ is the Boolean constant \textit{True}, $\lnot$ and $\land$ are the Boolean \textit{negation} and  \textit{conjunction} operators, respectively. 
$\mu$ is a \textit{predicate} over signals of the form $\mu:=l(x_t)\geq 0$, where $l: \mathbb{R}^n \to \mathbb{R}$ is a Lipschitz continuous function. $I=[a,b]=\{t\in\mathbb Z_{\geq0}\ |\ a\leq t\leq b;\ a,b\in \mathbb Z_{\geq0}\}$ denotes a bounded time interval and $\mathbf{U}$ is the temporal \textit{until} operator. 
The semantics of STL formulas is defined over signals. 
$\varphi_1 \mathbf{U}_{I} \varphi_2$ states that \enquote{$\varphi_2$ becomes true at some time point within $I$ and $\varphi_1$ must be always true prior to that.} The Boolean constant $\bot$ (\textit{False}) and \textit{disjunction} $\lor$ can be defined from $\top$, $\lnot$, and $\land$ in the usual way.
Additional temporal operators, \textit{eventually} and \textit{always}, are defined as $\mathbf{F}_I\varphi := \top\mathbf{U}_I\varphi$ and $\mathbf{G}_I\varphi := \neg\mathbf{F}_I\neg\varphi$, respectively. 
$\mathbf{F}_{I}\varphi$ is satisfied if \enquote{$\varphi$ becomes \textit{True} at some time in $I$} while $\mathbf{G}_{I}\varphi$ is satisfied if \enquote{$\varphi$ is \textit{True} at all times in $I$}.

The STL {\em qualitative semantics} \cite{maler2004monitoring} determines \textit{whether} a signal $X$ satisfies a formula $\varphi$ (written as $X\models\varphi$) or not (written as $X\not\models\varphi$). Its {\em quantitative semantics}, or \textit{robustness}, assigns a real value to measure \textit{how much} a signal satisfies $\varphi$. For instance, let $\varphi=\mathbf F_{[0,2]}x\geq2$. A signal $X_1=0,2,4,\ldots$ should have a higher robustness than $X_2=0,1,2.01,\ldots$ because $X_1$ satisfies $\varphi$ stronger. Multiple functionals have been proposed to capture the STL robustness \cite{donze2010robust,haghighi2019control,varnai2020robustness,gilpin2020smooth}. In this paper, we use the robustness function proposed in \cite{varnai2020robustness}, which is a \emph{sound} score, i.e., positive robustness indicates satisfaction of the formula, and negative robustness indicates a violation. As opposed to the traditional robustness \cite{donze2010robust}, the robustness in \cite{varnai2020robustness} is differentiable
so it is suitable for gradient-based optimization methods.  We denote the robustness of $\varphi$ at time $t$ with respect to signal $X$ by $\rho(\varphi,X,t)$. For brevity, we denote $\rho(\varphi,X,0)$ by $\rho(\varphi,X)$. The time horizon of an STL formula $\varphi$ denoted by $hrz(\varphi)$ is the closest time point in the future for which signal values are needed to compute the robustness at the current time \cite{dokhanchi2014line}. 

\subsection{Recurrent Neural Network (RNN) Controller}
Due to the \emph{history-dependence} property of STL, a feedback controller for system (\ref{eq:dynamics})
required to satisfy an STL formula over its state at time $t$, should be a function of the current state and all the history states, i.e., $u_t=\pi(x_{0:t})$. In this paper, we will learn a controller in the form of an RNN \cite{goodfellow2016deep}, which can be formulated as follows: 
\begin{equation}
    \label{eq:RNN}
    \begin{aligned}
    \mathbf h_t &=\mathcal{R}(x_t,\mathbf h_{t-1}, W_1),\\
    u_t &= \mathcal{N}(\mathbf h_t, W_2),
    \end{aligned}
\end{equation}
where $\mathbf h_t$ is the RNN hidden state, $t=0,1,\ldots,T-1$ ($T$ is the final time of interest)
At time $t$, the input of the RNN is the current state $x_t$ and the output is the control  $u_t$ for system (\ref{eq:dynamics}). $\mathcal R$ represents fully connected layers with parameters $W_1$, which update the hidden state based on the input and the previous hidden state. $\mathcal N$ denotes layers with parameters $W_2$, which predict the output based on the hidden state. The output layer is applied a hyperbolic tangent function to satisfy the constraint $u\in\mathcal U$ as in \cite{yaghoubi2020worst}. The hidden state $\mathbf h_t$ encodes all the previous inputs. $\mathbf h_{-1}$ is a zero vector.

\subsection{Discrete Time Control Barrier Function}
Consider system \eqref{eq:dynamics} and let $b:\mathbb R^n \rightarrow \mathbb R$. The set $\boldsymbol C=\{x\in\mathbb R^n\ |\ b(x)\geq0\}$ 
is (forward) {\em invariant} for system 
\eqref{eq:dynamics} if all its trajectories remain in $\boldsymbol C$ for all times, if they originate in $\boldsymbol C$. Function $b$ is a (discrete-time, exponential) {\em Control Barrier Function} (CBF) \cite{agrawal2017discrete} for system \eqref{eq:dynamics} if there exists $\alpha\in[0,1]$,  and for each $x_t$ there exists $u_t\in\mathcal U$ such that:
\begin{equation}\label{eq:cbf}
\begin{array}{c}
    b(x_0)\geq0\\
    b(x_{t+1}) + (\alpha-1)b(x_t)\geq0,\quad \forall t\in\mathbb Z_{\geq0},
    \end{array}
\end{equation}
where $x_{t+1}$, $x_t$, and $u_t$ are related by \eqref{eq:dynamics}. The set $\boldsymbol C$ is invariant for system \eqref{eq:dynamics} if there exists a CBF $b$ as in \eqref{eq:cbf} \cite{agrawal2017discrete}. This invariance property is usually referred to as {\em safety}. In other words, the system is safe if it stays inside the set $\boldsymbol C$.

\section{Problem Statement and Approach}

Consider system \eqref{eq:dynamics} with unknown dynamics $f$, $g$ and fully observable state. We assume that the initial state $x_0$ is randomly located in a set $\mathcal X_0\subseteq\mathcal X$ with the probability density function $p:\mathcal X_0\rightarrow\mathbb R$. Consider an STL formula $\varphi=\varphi_{task}\land\varphi_{safe}$ with predicates interpreted over $x$, where $\varphi_{task}$ is a general full STL formula that specifies the task, while $\varphi_{safe}=\mathbf{G}_{I_1}\phi_1\land\ldots\land\mathbf{G}_{I_N}\phi_{N}$ represents $N$ safety requirements. We assume that $\phi_1,\ldots,\phi_N$ are affine or quadratic predicates, so they can be converted into $N$ affine or quadratic CBFs $b_1,\ldots,b_N$. For example, the satisfaction of $\phi=(a^\top x_t \leq c)$ is equivalent to $b=-a^\top x_t + c\geq0$, where $a\in\mathbb R^n$ and $c\in \mathbb{R}$. We also assume that all CBFs have a relative degree of $1$. High-order CBFs will be investigated in future work. Given a finite time horizon $T\geq hrz(\varphi)$ (for simplicity we assume that $T=hrz(\varphi)$), our goal is to find a control policy $u_t=\pi(x_{0:t}),\ t=0,\ldots,T-1$ that makes trajectories $x_{0:T}$ starting from $\mathcal X_0$ satisfy $\varphi$. Since the robustness is sound, we can use it as a reward, and find a controller that maximize its expectation. 

In this paper we use a parameterized policy, which is implemented by an RNN. The control $u_t$ is computed by using \eqref{eq:RNN} recursively. We denote $u_t=\pi(x_{0:t};W)$, where $W = (W_1,W_2)$ captures the RNN parameters. We now formally state the STL control synthesis problem as follows:

\begin{Problem}
\label{pb:whole}
Given system \eqref{eq:dynamics} with unknown dynamics $f$ and $g$, fully observable state, initial state $x_0$ with distribution
$p:\mathcal X_0\rightarrow\mathbb R$, and given an STL formula $\varphi$ over $x$ with $T=hrz(\varphi)$, find the optimal policy parameters $W^*$ that maximize the expected STL robustness, i.e., 
\vspace{-2pt}
\begin{equation}
    \label{eq:whole}
    \begin{aligned}
    &W^*=\arg\max_{W}E_{p(x_0)} [\rho(\varphi,x_{0:T})]\\
    &\textrm{s.t.}~ x_{t+1} = f(x_t) + g(x_t)\pi(x_{0:t}; W), t=0,\ldots,T-1
    \end{aligned}
\vspace{-3pt}
\end{equation}
\end{Problem}

\begin{figure}
\vspace{+5pt}
  \centering
  \includegraphics[height=5cm]{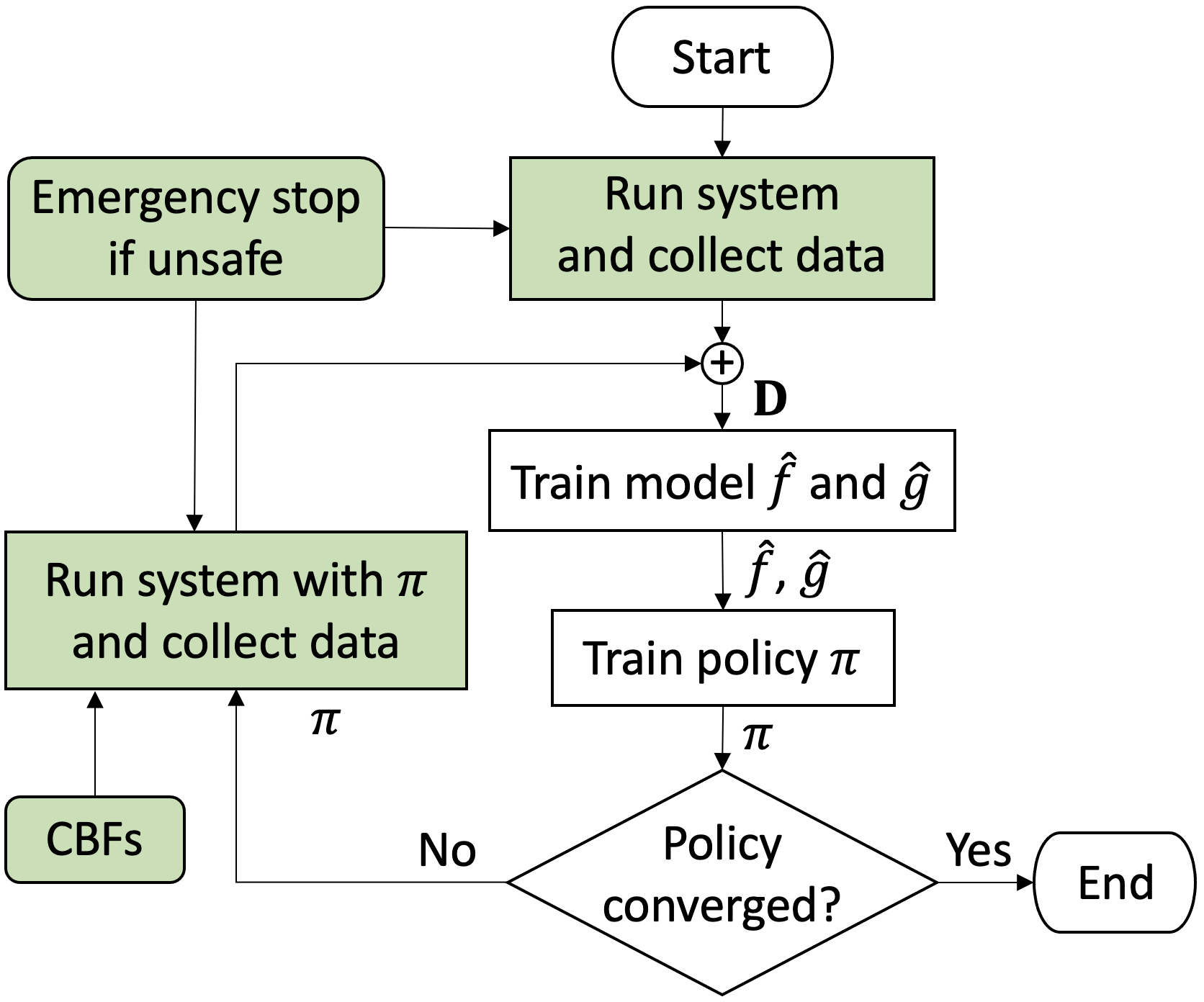}
  \caption {\small Our overall approach to Pb. \ref{pb:whole}: We first generate an initial dataset $\mathbf D$, and then we train the system model and the control policy alternately. Safety is improved by CBFs and ensured by emergency stops. The iteration terminates after convergence.}
  \label{fig:overview}
  \vspace{-5pt}
\end{figure}

Since the system model is unknown, Pb. \ref{pb:whole} cannot be solved directly. We first execute the system with random controls and collect the resulting system state transitions to form a dataset. We train two FNNs on this dataset to approximate the model (one for $f$ and one for $g$). Then we solve Pb. \ref{pb:whole} using the learned model to improve the control policy. We alternately improve the model using the data collected by applying the new policy, and improve the policy using the new model until convergence. CBFs are applied at each time the system is executed to improve safety except for generating the initial dataset. We stop the system and terminate this trial if
it gets too close to unsafe states. The overall model-based safe policy search framework is shown in Fig. \ref{fig:overview}. The technical details are presented in Sec. \ref{sec:solution}. 

Model-based learning is much more data-efficient than model-free RL methods, i.e., it requires fewer trials to learn the policy \cite{deisenroth2011pilco}. Alternately training the system model and the control policy benefits both. With a better policy, more states that the system may reach in order to finish the task can be explored, which results in a better model. With a more precise model, the policy can be better tuned. On the other hand, using CBF can not only improve safety but also make data collection more efficient, because the system can evolve for more steps in each trial without emergency stops.

\section{Model-Based safe policy search}
\label{sec:solution}

In this section, we introduce the model-based safe policy search approach. The system model learning, the control policy improvement, and the overall algorithm are presented in Secs. \ref{sbsec:Model}, \ref{sbsec:policy} and \ref{sbsec:whole}, respectively. 

\subsection{System Model Learning}
\label{sbsec:Model}
Although the true system dynamics \eqref{eq:dynamics} are assumed to be deterministic, learning a deterministic model to train the policy suffers from model bias, because an inaccurate model is given full confidence \cite{deisenroth2011pilco}. Hence, we use a probabilistic model implemented by two feedforward neural networks (FNNs) $\mathcal F$ and $\mathcal G$ with dropout to approximate the dynamics \eqref{eq:dynamics} given the current state $x_t$ and the control input $u_t$:
\vspace{-2pt}
\begin{equation}
    \label{eq:model}
    \hat x_{t+1} = \mathcal F(x_t;W_f,Z_f) + \mathcal G(x_t;W_g,Z_g)u_t
    \vspace{-2pt}
\end{equation}
where $W_f$ and $W_g$ are FNN parameters, $Z_f$ and $Z_g$ are random dropout masks that randomly deactivate the nodes in the dropout layers with probability $p_{d}$. As an approximation of a Bayesian Neural Network, an FNN with dropout can not only reduce model bias (detailed in Sec. \ref{sbsec:policy}), but also represents the uncertainty of the learned model via moment-matching \cite{gal2016dropout}. To do this, we evaluate the FNNs with the input $(x_t,u_t)$ and randomly sampled dropout masks $(Z_f^1,Z_g^1)\ldots,(Z_f^{N_z},Z_g^{N_z})$, and obtain the outputs $\hat x_{t+1}^1,\ldots,\hat x_{t+1}^{N_z}$. The model uncertainty at $(x_t,u_t)$ is represented by the covariance $\Sigma_t$ of $\hat x_{t+1}^1,\ldots,\hat x_{t+1}^{N_z}$. The FNNs can be easily transformed into a deterministic form by applying masks $Z_f^{det}$ and $Z_g^{det}$ that activate all nodes and scaling the outputs of dropout layers by a factor of $1-p_{d}$.

To learn the model, we need to execute the system and record the system transitions. The model learning algorithm is summarized in Alg. \ref{alg:1} (values in brackets are for initial dataset generation). At the beginning, we sample $N_0$ random initial states $\{x_0^1,\ldots,x_0^{N_0}\}$ from the probability distribution $p$ (step 2). From each initial state, we apply random controls $u_t\sim\mathrm{Uniform}(\mathcal U)$ and collect all the system transitions $\big((x_t^i,u_t^i),x_{t+1}^i\big)$ with $i=1,\ldots,N_0$ and $t=0,\ldots,T_0$ to create the initial dataset $\mathbf D$, where $T_0\geq T$ is the time horizon to apply random controls (step 5). Then two FNNs are trained on $\mathbf D$ simultaneously (step 11) to minimize the loss $C$:
\vspace{-2pt}
\begin{equation}
    \label{eq:train_model}
    C=\sum_{\mathbf D}\sum_{t=0}^{T-1}\|x_{t+1}-\hat x_{t+1}\|^2.
    \vspace{-2pt}
\end{equation}
After training, the FNN parameters are denoted by $W_f^*$, $W_g^*$.

Similarly to generating the initial dataset, each time after the control policy is improved, we sample $N_s$ initial states (step 2), starting from which the safe controls modified from the control policy $\pi$ using CBFs (discussed later) are applied until arriving at the time horizon $T$ (steps 7-8). We add all the system transition data to the dataset $\mathbf D$. Then the FNNs are retrained on the new dataset to minimize the loss $C$ (step 11). In practice, during all the system executions, if one of the values of $b_1,\ldots,b_N$ is less than $0$, which means that the CBFs fail due to model error, we stop the system and terminate the trial to protect the system (steps 9-10). 

As mentioned earlier, after the initial training of the model, we use CBFs with the learned model to modify the control policy. Here we transform the FNN model \eqref{eq:model} into the form:
\vspace{-2pt}
\begin{equation}
    \label{eq:cbf-model}
    \hat x_{t+1} = \mathcal F(x_t;W_f,Z_f^{det}) + \mathcal G(x_t;W_g,Z_g^{det})u_t+\delta_t,
    \vspace{-2pt}
\end{equation}
where $\delta_t$ is a zero-mean random vector representing the uncertainty of the learned model. Its covariance $\Sigma_t$ is estimated via moment-matching. For each CBF $b_j$, $j=1,\ldots,N$, the uncertainty of $x_{t+1}$ is passed to $b_j(x_{t+1})$ as an error $\epsilon_{t,j}$:
\vspace{-2pt}
\begin{equation}
    \label{eq:cbf-error}
    b_j(x_{t+1}) = b_j(\hat x_{t+1}) + \epsilon_{t,j}.
    \vspace{-2pt}
\end{equation}
We use the first order derivative to approximate the error:
\vspace{-2pt}
\begin{equation}
    \label{eq:error}
    \epsilon_{t,j} = \frac{\partial b_j}{\partial x}\Big|_{x=\mathcal F(x_t;W_f,Z_f^{det}) + \mathcal G(x_t;W_g,Z_g^{det})u_t}\cdot\delta_t.
    \vspace{-2pt}
\end{equation}
For brevity, in what follows, we omit the point that the derivative is evaluated at. The error $\epsilon_{t,j}$ is a zero-mean random variable with variance $\sigma_{t,j}^2$, which is given by:
\vspace{-2pt}
\begin{equation}
\label{eq:variance}
    \sigma_{t,j}^2 = (\frac{\partial b_j}{\partial x})\Sigma_t(\frac{\partial b_j}{\partial x})^\top.
    \vspace{-2pt}
\end{equation}
By choosing a lower limit $-\lambda\sigma_{t,j}$ of the error $\epsilon_{t,j}$ in \eqref{eq:cbf-error} to compute $b_j(x_{t+1})$, $\lambda>0$, we can guarantee the safety for all $\epsilon_{t,j}\geq-\lambda\sigma_{t,j}$. At time $t$ and state $x_t$, we obtain the modified control $u_t^{s}$ that is safe for errors above $-\lambda\sigma_{t,j}$:
\begin{equation}
    \label{eq:cbf-opt}
    \begin{aligned}
    u_t^{s} =& \arg\min_{u_t}\ \|u_t-\pi(x_{0:t},W)\|^2\\
    \textrm{s.t.}\quad &b_j\big(\mathcal F(x_t;W_f,Z_f^{det}) + \mathcal G(x_t;W_g,Z_g^{det})u_t\big) \\
    & + (\alpha-1)b_j(x_t) \geq \lambda\sigma_{t,j},\quad j=1,\ldots,N
    \end{aligned}
    \vspace{-1pt}
\end{equation}
where $\alpha\in[0,1]$. In \eqref{eq:cbf-opt}, $\sigma_{t,j}$ is supposed to be a function of $u_t$ (the derivative ${\partial b_j}/{\partial x}$ depends on $u_t$), but we assume a constant $\sigma_{t,j}$ that is evaluated at $u_t=\pi(x_{0:t},W)$ to make \eqref{eq:cbf-opt} simpler.  Since $b_j$ is affine or quadratic, \eqref{eq:cbf-opt} is a (quadratically constrained) quadratic program, which can be solved efficiently using solvers such as Gurobi \cite{gurobi}. 

\begin{Remark}
The safe control $u_t^s$ can only improve safety, rather than fully guarantee it, because: (1) estimate bias exists; (2) the error $\epsilon_{t,j}$ in $b_j(x_{t+1})$ is approximated by the first order derivative and with a fixed control $\pi(x_{0:t},W)$; (3) the error may exceed the lower limit. However, according to our simulation results (Section \ref{sec:simulation}), this method can largely reduce the chance of the system reaching unsafe regions. 
\end{Remark}

\vspace{-10pt}
\begin{algorithm}
\KwIn{A control policy $\pi$, a dataset $\mathbf D$}
\KwOut{Optimal FNN parameters $W_f^*$, $W_g^*$}
\For{$i\in1,\ldots,N_s(N_0)$}{
Sample initial state $x_0^i$\;
\For{$t=1,2,\ldots,T (T_0)$}
{\eIf{No policy and model was trained}{Apply $u_t^i\sim\mathrm{Uniform}(\mathcal U)$ to the system and add the data $\big((x_t^i,u_t^i),x_{t+1}^i\big)$ to $\mathbf D$\;}
{Adjust $\pi(x_{0:t}^i,W)$ to obtain $u_t^{s,i}$ by \eqref{eq:cbf-opt}\;
Apply $u_t^{s,i}$ to the system and add the data $\big((x_t^i,u_t^{s,i}),x_{t+1}^i\big)$ to $\mathbf D$\;}
\If{$\min(b1,\ldots,b_N)\leq 0$}{Break}}}

Train the FNN on $\mathbf D$, \Return $W_0^*$\;
 \caption{System model learning}\label{alg:1}
\end{algorithm}
\vspace{-15pt}

\subsection{Control Policy Improvement}
\label{sbsec:policy}
Now we can solve \eqref{eq:whole} in Pb.~ \ref{pb:whole}. We substitute the unknown system dynamics with the learned one from Sec.~\ref{sbsec:Model}. 
Also, we estimate the expected robustness over initial conditions by sampling  $M$ initial states $x_0^1,\ldots,x_0^M$ and averaging the resulting robustness values. For different initial states, different dropout masks $(Z_f^1,Z_g^1)\ldots,(Z_f^{M},Z_g^{M})$ in the learned model are sampled to estimate the trajectories (no CBFs applied). The average over initial states is also an average over system models, hence it can reduce model-bias \cite{deisenroth2011pilco}. For brevity, let $\hat{f}^i(x_t) = \mathcal F(x_t;W_f,Z_f^i)$ and $\hat{g}^i(x_t) = \mathcal G(x_t;W_g,Z_g^i)$ be the sampled system model. Now the problem becomes:
\vspace{-2pt}
\begin{equation}
    \label{eq:policy_update}
    \begin{aligned}
    W^*=&\arg\max_{W}\frac{1}{M}\sum_{i=1}^M \rho(\varphi,x_{0:T}^i)\\
    \textrm{s.t.}\quad &x_{t+1}^i = \hat f^i(x_t^i) + \hat g^i(x_t^i)\pi(x_{0:t}^i,W),\\
    &t=0,1,\ldots,T-1,\ i=1,\ldots,M.
    \end{aligned}
\end{equation}
We solve \eqref{eq:policy_update} by substituting the constraints into the objective function and solving the unconstrained optimization problem. We use gradient ascent to iteratively update the RNN parameters $W$. At each iteration step, we randomly re-sample a new set of $M$ initial states and $M$ dropout masks to evaluate the objective function and its gradient. Doing so gives us an unbiased estimator of the objective \cite{gal2016improving}. We then implement the optimization using the stochastic optimizer Adam \cite{kingma2014adam}. The optimization terminates after convergence. The policy improvement process is shown in Alg. \ref{alg:2}.

\vspace{-10pt}
\begin{algorithm}
\SetAlgoLined
\KwIn{Learned model $\hat f$, $\hat g$, an STL formula $\varphi$}
\KwOut{Optimal policy parameters $W^*$}
\Repeat{Convergence; \Return $W^*$}{
Sample initial states $\{x_0^1,\ldots,x_0^M\}$\; 
Sample system models $\{(\hat f^1,\hat g^1),\ldots,(\hat f^M,\hat g^M)\}$\;
Compute the objective in \eqref{eq:policy_update}\;
Back-propagate the gradient $\Delta W$ of \eqref{eq:policy_update}\;
Update $W$ with $\Delta W$ using the Adam optimizer\;
}
\caption{Control policy improvement}\label{alg:2}
\end{algorithm}
\vspace{-10pt}

\begin{Remark}
All the derivatives involving the FNN system model and the RNN controller can be computed using the auto-differentiation tools designed for neural networks (e.g., PyTorch \cite{paszke2017automatic}). When computing the STL robustness $\rho$, we applied the method STLCG \cite{leung2020back} which uses a computation graph similar to a neural network. So the auto-differentiation tools can also be applied. Hence, the gradient of the objective function $\frac{\partial\rho}{\partial W}$ in \eqref{eq:policy_update} can be accessed easily and analytically.
\end{Remark}

\subsection{Model-based policy search}
\label{sbsec:whole}
We alternately use Alg. \ref{alg:1} and Alg. \ref{alg:2} to train the model and the policy as in Fig. \ref{fig:overview}. We call the completion of one model learning and one policy improvement a \emph{cycle}.  The algorithm will terminate and return the final control policy with parameter $W^*$ when the policy converges. 

\begin{Remark}
Although the objective of our algorithm is to maximize the expected STL robustness over trajectories starting from random initial states, there is no guarantee that the learned policy can satisfy the STL from all initial states. Moreover, using CBFs might change the original control given by the RNN, which might decrease the robustness or even violate the STL specification. This also causes that in model learning phase the robot explores different regions from those induced by the learned policy. Hence, the model may not be properly learned in some regions explored by the learned policy. However, by also including the safety constraints into the STL formula, the RNN controller is aware of these constraints, which avoids dramatic changes when using CBFs. This mitigates the above problems. It is also possible to include CBFs in the policy learning phase by differentiating through \eqref{eq:cbf-opt} using technique in \cite{amos2017optnet}. We will investigate this in future work. The simulation and experiment results show that our approach can reach a very high success rate of satisfying the STL specification. 
\end{Remark}

\begin{Remark}
The advantages of using neural networks (RNN and FNN) includes: (1) fast execution; (2) good scalability; (3) easy propagation of gradients. These benefits will be shown in Sec. \ref{sec:simulation}. An additional advantage of the RNN is the ability to satisfy general full STL formulae.  Another choice is using a memoryless controller but augmenting the system state \cite{aksaray2016q}, \cite{venkataraman2020tractable}, which makes the state space too large for long horizon control and is restricted to STL fractions. 
\end{Remark}

\section{Simulations}
\label{sec:simulation}

\begin{figure}
\vspace{+5pt}
     \centering
     \begin{subfigure}[b]{0.23\textwidth}
         \centering
         \includegraphics[height=3.4cm]{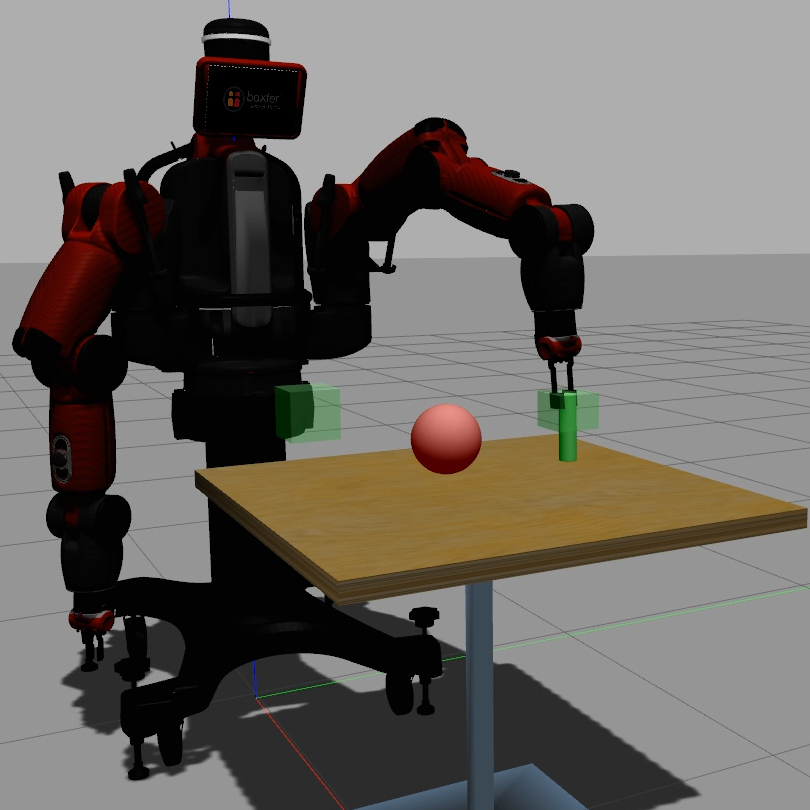}
         \caption{\small Pick}
     \end{subfigure}
     \begin{subfigure}[b]{0.23\textwidth}
         \centering
         \includegraphics[height=3.4cm]{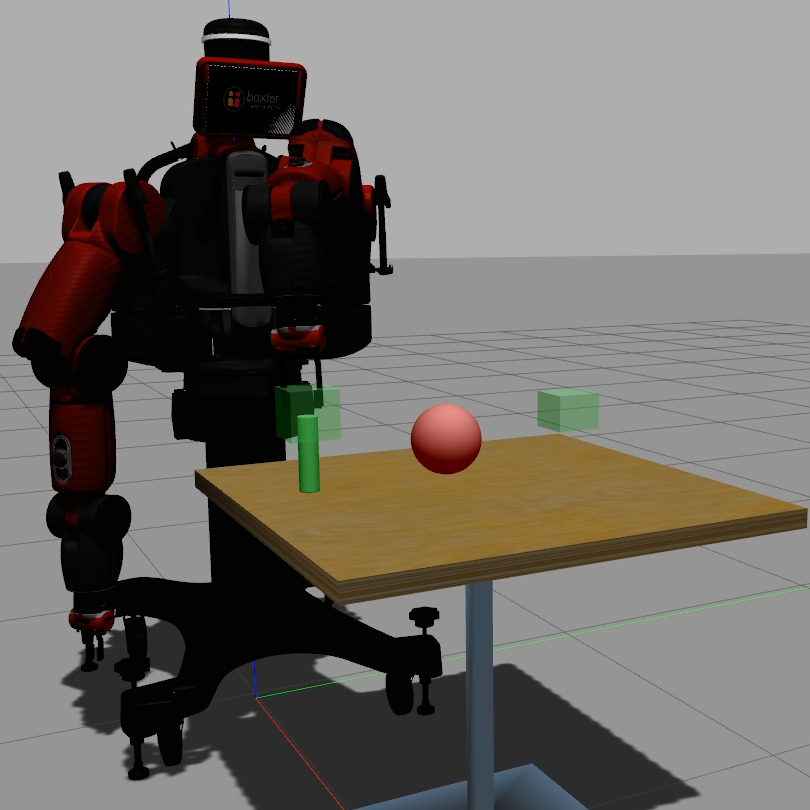}
         \caption{\small Place}
     \end{subfigure}
     \vspace{-6pt}
    \caption{\small The Gazebo simulation environment. The green cylinder is the object to be picked, and the red ball is the obstacle. The red, green, blue lines are the $x$, $y$, $z$ axes respectively. Transparent boxes are $R_{pick}$ and $R_{place}$. Video is at https://youtu.be/UmuJdVcxY14.}
\label{fig:baxter}
\vspace{-8pt}
\end{figure}

 We consider a pick-and-place task for a Baxter robot in Gazebo \cite{koenig2004design} simulation environment (shown in Fig. \ref{fig:baxter}).  The algorithms were implemented in Python. The NNs were implemented using Pytorch \cite{paszke2017automatic}. We used a computer with a 3.50GHz Core i7 CPU and 16GB RAM for the simulation.

Baxter is a $7$ DOF dual arm robot. We only use its left arm in this case study. The system state $x_t\in\mathbb R^{14}$ is defined as the concatenation of joint angles $\theta_t\in\mathbb R^7$ and joint velocities $\dot\theta_t\in\mathbb R^7$. We use the torques applied to the joints as the control input $u_t\in [-8,8]^4\times[-3,3]^3 \subseteq \mathbb R^7$. Note that $u_t$ is applied in addition to the known gravity compensation torques. We control the system with a time interval $\Delta t = 1/6s$. The initial joint angles are $[-0.6,-0.75,0,0,75,0,1.46,0]$ rad with a uniformly sampled disturbance $d\in[-0.1,0.1]^7$. 

Our specification is defined over the end effector pose $[\mathbf p_t,\mathbf q_t]$ where $\mathbf p_t=[p_x,p_y,p_z]$ is the 3D position and $\mathbf q_t=[q_w,q_x,q_y,q_z]$ is the quaternion representing the orientation. Assume the forward kinematics $[\mathbf p_t,\mathbf q_t]=K(\theta_t)$ are known and the motion primitives for \emph{open} and \emph{close} of the gripper are already available. Both \emph{open} and \emph{close} can be finished within $0.5s$. Then the STL task is: stay in the \emph{pick} region ($R_{pick}$) for $0.5s$ between $[0s,3s]$, and stay in the \emph{place} region ($R_{place}$) for $0.5s$ between $[3s,6s]$. We also require the rotation angle of the end-effector from the vertical pose be always less than $30^\circ$ ($\mathbf q\in V_{30}$) so that the gripper is downward when picking up the object. The safety constraints are to always avoid collision with a spherical obstacle ($\mathbf p\not\in Obs$), 
its body ($p_x \geq x_{body}$), and the table ($p_z\geq h_{tb}$). The task can be formulated as $\varphi_1$ with $hrz(\varphi_1)=36$:
\vspace{-3pt}
\begin{equation*}
\begin{aligned}
    \varphi_1 = &\big(\mathbf F_{[0,\frac{2.5}{\Delta t}]}(\mathbf G_{[0,\frac{0.5}{\Delta t}]} \mathbf p\in R_{pick})\big) \\
    \land & \big(\mathbf F_{[\frac{3}{\Delta t},\frac{5.5}{\Delta t}]} (\mathbf G_{[0,\frac{0.5}{\Delta t}]} \mathbf p\in R_{place})\big) \land (\mathbf G_{[0,\frac{6}{\Delta t}]} \mathbf q\in V_{30})  \\
    \land & (\mathbf G_{[0,\frac{6}{\Delta t}]} p_x>x_{body}) \land (\mathbf G_{[0,\frac{6}{\Delta t}]} p_z>h_{tb})  \\ \land & (\mathbf G_{[0,\frac{6}{\Delta t}]} \mathbf{p}\not\in Obs)
\end{aligned}
\vspace{-3pt}
\end{equation*}
Although the CBFs representing the safety constraints are affine or quadratic with respect to the end-effector position, they are nonquadratic with respect to the system state due to the complexity of the forward kinematics. Hence, we linearize the forward kinematics at each time step, which makes the CBFs meet the affine or quadratic assumption.

Let $N_0=100, N_s = 20, M=200$, $N_z=20$, $\lambda=3$, $\alpha=0.8$. The RNN applies an LSTM \cite{hochreiter1997long} with $3$ hidden layers each with $64$ nodes. Both FNNs have $2$ hidden layers each with $128$ nodes and the dropout probability $p_{d}=0.05$.  After $5$ cycles of training which takes $50$min, the robot can finish the task successfully as shown in Fig. \ref{fig:baxter}. This is much faster than our previous work~\cite{liu2021recurrent} in which training the controller for a much easier task takes $2$ hours.
We test the policy (trained with CBFs) after each cycle on $100$ random initial states. The success rate $\gamma$ (getting positive robustness), unsafe rate $\beta$ (violating safety constraints) and average robustness $\bar\rho$, with or without CBFs, are shown in Table \ref{tb:1}. 

\begin{table}
\centering
\caption{\small Success Rate and Average Robustness for $\varphi_1$}
\label{tb:1}
\begin{tabular}{l|ccc|ccc}
\hline
& \multicolumn{3}{c}{w/ CBF} & \multicolumn{3}{|c}{w/o CBF} \\
\cline{2-7}
& $\gamma$ & $\beta$ & $\bar\rho$ & $\gamma$ & $\beta$ & $\bar\rho$\\
\hline
Cycle 1 & $0\%$ & $10\%$ & -0.4807 & $0\%$ & $24\%$ & -0.7224\\
\hline
Cycle 2 & $2\%$ & $0\%$ & -0.0973& $14\%$ & $5\%$ & -0.0880\\
\hline
Cycle 3 & $14\%$ & $0\%$ & -0.0235& $22\%$ & $0\%$ & -0.0071\\
\hline
Cycle 4 & $83\%$ & $0\%$ & 0.0169& $87\%$ & $0\%$ & 0.0192\\
\hline
Cycle 5 & $100\%$ & $0\%$ & 0.0401& $100\%$ & $0\%$ & 0.0419\\
\hline
\end{tabular}
\vspace{-5pt}
\end{table}

After $5$ cycles of training the success rate reaches $100\%$. From Table \ref{tb:1}, we can see that after updating the model, the policy can be improved to get a higher robustness and a higher success rate. The unsafe rate $\beta$ is largely reduced by using CBFs. Note that sometimes the robustness and success rate without CBFs are higher. This is because CBFs can change the policy. The system runs for a total of $180$ times, which is a relatively small number. In \cite{aksaray2016q} a model-free Q-learning approach runs $2000$ episodes to learn the policy for a much easier STL task and a much simpler system. After training, execution of the RNN controller and solving \eqref{eq:cbf-opt} to obtain the safe control take $0.0004s$ and $0.03s$ (in average) respectively. When the discrete time interval is large enough compared with the computation time for the safe control, the controller can be implemented in real time. The results show that the algorithm proposed in this paper is capable to control a robot arm directly from joint torques to end-effector pose without knowing the complex system dynamics. The planning horizon, system dimension and number of predicates would affect the training time linearly, while they have little effect on the execution time after training, hence our approach has a good scalabilty.

\section{Experiment}
\begin{figure}
\vspace{+5pt}
  \centering
  \includegraphics[height=3.5cm]{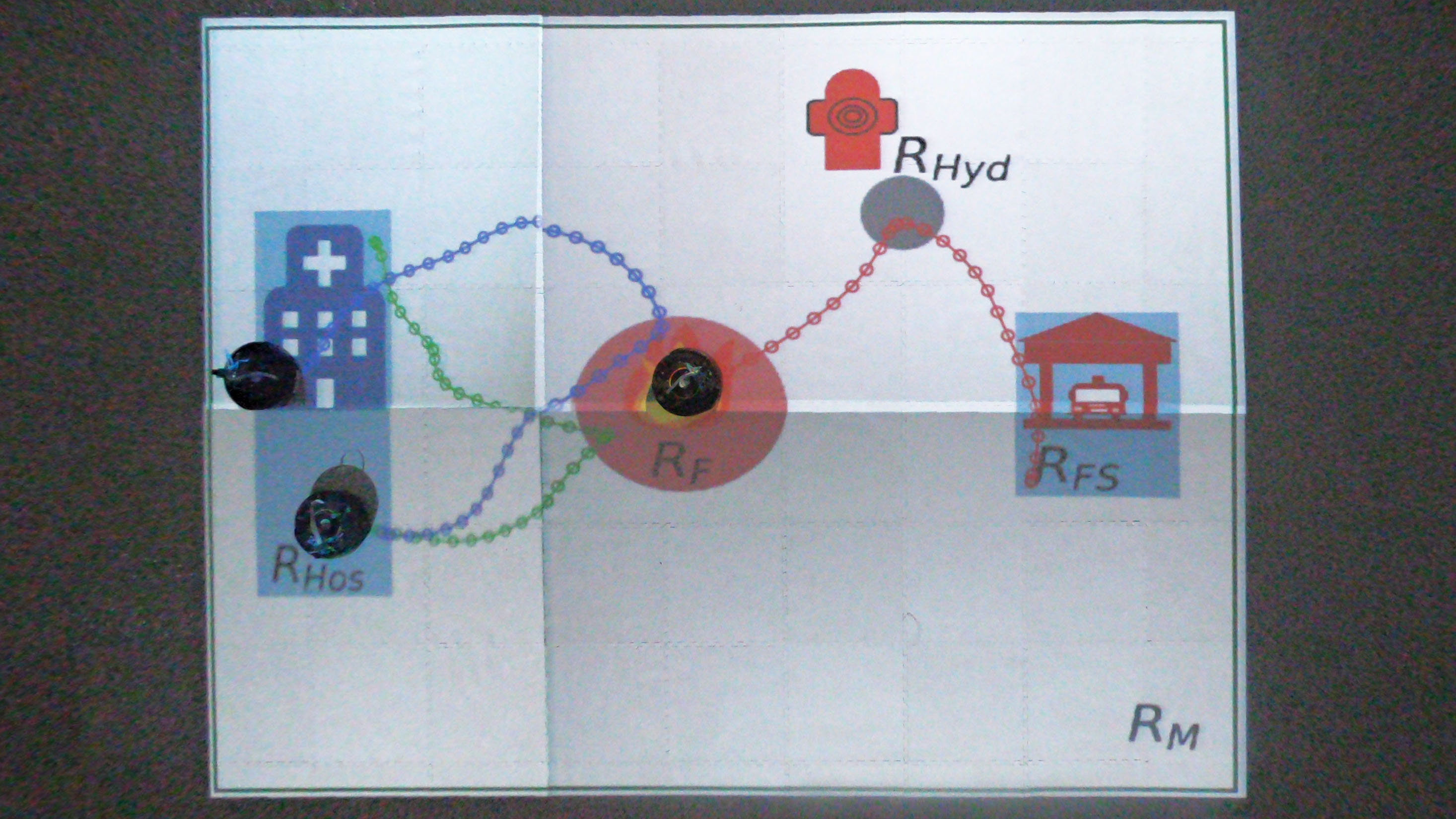}
  \caption {\small An execution of the system with the learned policy. The trajectories of the robots are projected on to the ground.}
  \label{fig:experiment}
  \vspace{-5pt}
\end{figure}

We consider a system containing $3$ iRobot Create2 ground robots in an indoor motion capture environment (as shown in Fig. \ref{fig:experiment}). The position and orientation of the robots can be obtained using the Optitrack motion capture system. Let $x^k=[p_{x}^k\ p_{y}^k\ \sin\theta^k\ \cos\theta^k]^\top$ and $u^k=[v^k\ \omega^k]^\top$ be the state and the control of the $k^{th}$ robot, where $p^k_x, p^k_y, \theta^k$ give the position and orientation and $u^k=[v^k\ \omega^k]^\top$ captures the speed and angular speed, with $v^k\in[0,0.75]$ and $\omega^k\in[-\frac{\pi}{2},\frac{\pi}{2}]$, $k\in\{1,2,3\}$. We normalize the terms corresponding to $\sin\theta$ and $\cos\theta$ in the output of the learned model after each step. The state and control of the whole system are the concatenations of the states and controls of the $3$ robots. Hence the system state $x\in\mathbb R^{12}$ and the control $u\in\mathbb R^6$. Controls are sent to the robots in a rate of $3 HZ$.  The algorithm is implemented on a computer with a 4.20GHz Core i7 CPU and 32GB RAM. All these platforms communicate using the Robot Operating System (ROS) \cite{ros}. 

Consider a fire-fighting scenario as shown in Fig. \ref{fig:experiment}, where two ambulances ($k=1,2$) randomly located in the hospital $R_{Hos}$ are supposed to go to the fire $R_{F}$ in $8s$, and then return to the hospital in $[8s,15s]$, while a fire engine ($k=3$) randomly located in a fire station $R_{FS}$ should go to the fire in $10s$ and stay there in $[10s,15s]$, and reaches a Hydrant $R_{Hyd}$ before that. The safety constraints are (1) never go out of bounds $R_M$; (2) the distances between the robots are always larger than $d$. Let $p^k=[p_{x}^k\ p_{y}^k]$. The specification can be formulated as $\varphi_2$ with $hrz(\varphi_2)=15/\Delta t = 45$:
\vspace{-3pt}
\begin{equation*}
    \begin{aligned}
    \varphi_2 = & \land_{k=1}^2\big(\mathbf{F}_{[0,\frac{8}{\Delta t}]}(p^k\in R_{F}) \land \mathbf{F}_{[\frac{8}{\Delta t},\frac{15}{\Delta t}]}(p^k\in R_{Hos}) \big) \\
    & \land \mathbf{F}_{[0,\frac{10}{\Delta t}]}(p^3\in R_{Hyd}) \land \mathbf{G}_{[\frac{10}{\Delta t},\frac{15}{\Delta t}]}(p^3\in R_F)\\
    &\land_{i,j\in\{1,2,3\},i\neq j}\mathbf{G}_{[0,\frac{15}{\Delta t}]}(\|p^i-p^j\|^2\geq d^2)\\
    & \land_{k=1}^3 \mathbf{G}_{[0,\frac{15}{\Delta t}]}(p^k\in R_M).
    \end{aligned}
    \vspace{-3pt}
\end{equation*}
Note that the safety predicates (and the corresponding CBFs) are all affine and quadratic functions.

Let $N_0=50$, $\alpha=0.5$. The other settings are same with the simulation. When computing the norm in \eqref{eq:cbf-opt} we assign a weight of $0.03$ to the angular speed in order to encourage the robot to turn instead of slowing down when approaching an obstacle. The computation time for the RNN policy and the safe control ($0.0004s$ and $0.01s$ respectively) is negligible compared with $\Delta t = 1/3s$, which enables real time control. The robots can learn a good policy that satisfies the specification after $4$ cycles of training. In this process, we collected $2372$ data points to train the system model, with the system running for about $13$ minutes (not counting the time it takes the robots to return to the initial states). The total training time is about $20$ minutes.  Fig. \ref{fig:experiment} shows an execution of the system with the learned policy. We tested the policy $20$ times and the success rate was $100\%$. 

\section{Conclusion}
In this paper, we proposed a model-based policy search approach to maximize STL robustness. The system model and the control policy are learned alternately. The policy is implemented as an RNN, which can deal with the history-dependence of STL. Simulation and experiment results show that our approach can learn the policy within relatively few system executions and achieves high success rate. Safety can be improved by using CBFs. After training, the RNN policy can be implemented in real time. 


\bibliographystyle{IEEEtran}
\typeout{} 
\bibliography{references}

\end{document}